\documentclass[11pt, A4]{article}
\usepackage{jheppub}

\newcommand{\be}{\begin{equation}}
\newcommand{\ee}{\end{equation}}
\newcommand{\p}{\partial}

\newcommand{\nn}{\nonumber}
\def\bea{\begin{eqnarray}}
\def\eea{\end{eqnarray}}
\def\m#1{\mathcal#1}
\def\al{{\alpha}}

\makeatletter \@addtoreset{equation}{section} \makeatother

\title{ Counterterms in Gravity in the Light-Front Formulation and 
a $D=2$ Conformal-like Symmetry in Gravity}
\author[a]{Anders K. H. Bengtsson,}
\author[b]{Lars Brink,}
\author[c]{Sung-Soo Kim}

\affiliation[a]{
School of Engineering,
University College of Bor\aa s\\
All\'{e}gatan 1, SE-50190 Bor\aa s, Sweden
}

\affiliation[b]{Department of Fundamental Physics,
Chalmers University of Technology,\\ S-412 96 G\"{o}teborg, Sweden}

\affiliation[c]{
School of Physics, Korea Institute for Advanced Study, \\
85 Hoegiro, Dongdaemun-gu, Seoul 130-722, Korea }
 
\emailAdd{Anders.Bengtsson@hb.se}
\emailAdd{lars.brink@chalmers.se}
\emailAdd{sungsoo.kim@kias.re.kr}

\abstract{In this paper we discuss gravity in the light-front formulation (light-cone gauge) and show how possible counterterms arise. We find that Poincar\'e invariance is not enough to find the three-point counterterms uniquely. Higher-spin fields can intrude and mimic three-point higher derivative gravity terms. To select the correct term we have to use the remaining reparametrization invariance that exists after the gauge choice. We finally sketch how the corresponding programme for $N=8$ Supergravity should work.
}
\arxivnumber{1212.2776}
\begin{document}
\begin{flushright}  
{\tt KIAS-P12080}
\end{flushright}
\maketitle


\section{Introduction}\label{sec:Introduction}
Einstein's gravity theory is perhaps the most beautiful theory ever constructed. It works over a fantastic range of scales. It is only when we approach Planck scales that we believe that it has to be augmented. As a quantum field theory it is obviously non-renormalizable, but if we only consider scattering amplitudes it works better than expected. This was first shown in the famous paper by 't Hooft and Veltman~\cite{tHooft:1974bx}, who showed that the S-matrix for pure gravity without matter indeed was finite at the one-loop level. After a remarkable effort Goroff and Sagnotti~\cite{Goroff:1985th}  finally showed that the two-loop on-shell amplitudes are infinite. From a particle physicist's point of view this means that the theory has to be modified in order to be a perturbatively finite theory. Both the Superstring Theory and the Supergravity theories are theories that avoid the problems of ordinary gravity at lower loop orders and at the same time open up to fundamental theories including all interactions. We have very strong indications that Superstring theories are indeed perturbatively finite~\cite{Mandelstam:1991tw, Berkovits:2004px}, and in the end this is where we should look for the Theory of Everything, but also the maximally supersymmetric ($N=8$) supergravity theory has shown remarkable quantum properties. For a review see~\cite{Bern:2011qn}.  There are strong indications that the theory is perturbatively finite up to the seven-loop level. This is based on both real hardcore numerical computations~\cite{Bern:2012uf} as well as the construction of possible counterterms~\cite{Howe:1980th, Bossard:2011tq, Bjornsson:2010wm, Bossard:2010bd, Beisert:2010jx}. However, as we will see in this paper the construction of counterterms is quite delicate and can never be the ultimate proof that a theory does diverge. In order to prove finiteness we do need an analytic proof of some kind. 

Two of the analytic proofs~\cite{Brink:1982wv, Mandelstam:1982cb} for the finiteness of  the sister theory the $N=4$ theory were based on light-front techniques (the light-cone gauge). We have previously also set up ordinary gravity~\cite{Bengtsson:1983pd} as well as $N=8$ Supergravity~\cite{Bengtsson:1983pg}  in this formalism. In this paper we will take that formalism further by asking what kind of counterterms can be constructed in this formalism. We do not expect, of course, to find anything but the already established results, but it is always interesting to view the problem from a different perspective. We will see in the course of the calculations that we must have a very precise knowledge of all the symmetries of the theory. The ``$lc_2$ formalism" of gravity~\cite{Bengtsson:1983pd} in which only the two physical degrees of freedom are present can be seen as a (non-linear) representation of the Poincar\'e algebra. This works in the construction of the gravity Lagrangian at least up to the four-point coupling. However, we will find that in order to find the appropriate counterterms we have to impose a further symmetry. This can be seen as the remnant of the reparametrization invariance once we have fixed the gauge to the light-cone gauge and eliminated all unphysical degrees of freedom. The remaining (residual) symmetry will look infinitesimally as a Virasoro-like symmetry in the transverse plane. Hence gravity in this formalism can be viewed to a certain extent as a $2d$ conformal theory imbedded in a $4d$ Poincar\'e invariant theory.

In section \ref{sec:LCformulations} we will set up gravity in the light-front formulation. We will do it as a purely algebraic exercise and show that it gives a unique three-point coupling. In section \ref{sec:counterterm} we will find possible counterterms by closing the Poincar\'e algebra. We will see that just finding a (non-linear) representation of the Poincar\'e algebra to that order will not determine the counterterms precisely. This is discussed in section \ref{sec:higherspin}, where the solution to this dilemma is given. It amounts to realize that there are contaminations from higher spins and we show that there is a remaining infinitesimal reparametrization invariance in the transverse plane left which fixes the counterterms uniquely. This symmetry contains indeed a Virasoro-like symmetry and in section \ref{sec:virasoro} we discuss its further use and end in section \ref{sec:n8sugra} with a discussion how to take the results from this paper over to the $N=8$ Supergravity theory.

\section{Gravity in the light-frame formulation}\label{sec:LCformulations}
In the light-frame formulation we start by introducing light-frame coordinates
\begin{align}
x^\pm= \frac{1}{\sqrt2}(x^0\pm  x^3), && 
\partial^{\pm}= \frac{1}{\sqrt2}(\partial^0\pm \partial^3),
\end{align}
satisfying $\partial^- x^+ =-1 = \partial^+ x^-$, 
and rewrite the transverse coordinates and its derivatives as complex entities
\begin{align}
x= \frac{1}{\sqrt2}(x^1+ix^2), && \bar x= \frac{1}{\sqrt2}(x^1-ix^2),\\
\partial= \frac{1}{\sqrt2}(\partial_1+i\partial_2),&&\bar\partial= \frac{1}{\sqrt2}(\partial_1-i\partial_2).
\end{align}
We use space-like metric ($\partial \bar x =\bar\partial x=1$).

We write the Poincar\'e generators $P^\mu$ and $J^{\mu \nu}$ with similar definitions and obvious notations as $P^+,P^-, P$ and $\bar P$, and $ J^+, \bar J^+, J^{+ -}, J,\, J^-$ and $\bar J^-$. Here $J = J^{12}$ is the helicity generator and measures the helicity of the field. According to Dirac we can take any direction within the light-cone as the time axis and we choose then $x^+$ to be the ``time". Since we are only going to look at massless fields we implement the mass-shell condition $P_\mu P^\mu =0$ and write for the free theory
\be
P^- = -i \frac{\partial \bar \partial}{\partial^+}.
\ee
The division with $\partial^+$ is quite harmless and will be done in most formulae in the sequel. It is a non-local operation which can be thought of as an integral. The relevant formula to remember is 

\be
\partial^+ \frac{1}{\partial^+} f(x^-) =  f(x^-).
\ee
One can also Fourier transform it and then it corresponds to a pole in the momentum $p^+$, and the burden is then to exactly define the pole. There are various descriptions for that~\cite{Mandelstam:1982cb, Leibbrandt:1983pj}. Since $P^-$ is conjugate to the time $x^+$ it is the Hamiltonian and $p^+$ can be thought of as the mass in a non-relativistic analogy.

Since $P^-$ is the Hamiltonian it will get non-linear terms in the interaction theory. We will implement the Poincar\'e generators by letting them act as derivative operators on the field. That will mean that the general form for the Hamiltonian will be
\be
P^- \varphi = \delta _{P^-} \varphi= -i \frac{\partial \bar \partial}{\partial^+}\varphi +D( \varphi , \varphi)+ F( \varphi , \bar\varphi) +  G( \varphi , \varphi, \bar\varphi)+\ldots\ . 
\ee
The equation of motion is found by setting
\be
-i \partial^- \varphi = P^- \varphi.
\ee

 All the generators with a ($-$) index such as $J^{+-}, J^-, \bar J^-$ will be generators that take the field forward in time. They will all have non-linear contributions. We will construct the algebra at $x^+=0$ which will mean that the generator $ J^{+ -}$ will still be linear. After Dirac we call the linearly realized generators the kinematical ones and the non-linearly realized ones the dynamical ones or the Hamiltonians.

When constructing the helicity generator one finds by closing the free algebra the most general form to be
\be
J = x \bar \partial - \bar x \partial - \lambda,
\ee
where $\lambda$ is an arbitrary number. When closing the algebra at the three-point level one finds it to be an integer, the helicity of the field. Here we only consider commuting fields.

Let us now summarize the result of paper~\cite{Bengtsson:1983pd}. For any even spin $\lambda$ one can construct a three-point self-interaction according to the formula
\be \label{H}
\delta^{\alpha}_{P^-} \varphi =  \al \sum_{n=0}^{\lambda}(-1)^n \binom{\lambda}{n}{\partial^+}^{(\lambda-1)}
\left [\frac{\bar \partial^{(\lambda -n)}}{{ \partial^+}^{(\lambda -n)}}\,\varphi \frac{{\bar \partial}^n}{{\partial^+}^n}\,\varphi\right] + F(\varphi, \bar{\varphi}) + G(\varphi, \varphi, \bar{\varphi})+ \ldots \,,
\ee 
where by $\delta^{\alpha}_{P^-}$ we denote the contribution of $\delta_{P^-}$ to (coupling constant) order $\alpha$.
 In the case of odd spin we get the same formulae but we have to introduce fields which carry a group index and we have to use a structure constant in the amplitude to get a proper antisymmetry. The factor $F(\varphi, \bar{\varphi})$ is determined by constructing the integral form corresponding to the variation (\ref{H}) in a second-quantized form. The term we have computed will be of the form $ \bar{\varphi}\,\, \varphi\,\, \varphi$. The term $F(\varphi, \bar\varphi)$ above will then come from the complex conjugate term of the integral form.  The Hamiltonian is real so we have to add the complex conjugate term. By taking the variation with respect to $\bar\varphi$ of that term will give us the term $F(\varphi, \bar\varphi)$. We will hence get that term for free in our computations. For completeness, we write 
\begin{align*}
F(\varphi, \bar\varphi)= 
\al \sum_{n=0}^{\lambda}(-1)^n \binom{\lambda}{n}\frac{2}{\partial^{+(\lambda-1)}}
\left [\frac{\partial^{(\lambda -n)}}{{ \partial^{+(\lambda -n)}}}\,\bar\varphi \,\,\partial^{+(2\lambda-n)}{\partial}^n\,\varphi\right] .
\end{align*} 
 
Let us now introduce a coherent state-like formalism \cite{Brink:2008qc} (See the Appendices for more details).  We construct the operators (partly for future use)
\begin{align}\label{E}
E=e^{a \hat{\bar \partial} + b\hat{\partial}}\qquad {\rm and}\qquad E^{-1}=e^{-a \hat{\bar \partial} - b\hat{\partial}},
\end{align}
with
\begin{align}
\hat{\bar \partial}=\frac{ \bar \partial}{\partial^+}\qquad {\rm and}\qquad \hat{\partial}=\frac{\partial}{\partial^+}.
\end{align} 
 We can then rewrite the Hamiltonian variation as 
\begin{align}
\delta^{\al}_{P^-} \varphi&=  \al \,\partial^{+(\lambda-1)} \left[\,e^{a \hat{\bar \partial}}  \varphi\, e^{-a \hat{\bar \partial}} \varphi \right] \Big|_{a^{\lambda}}+ F(\varphi, \bar\varphi)+ ...,
\end{align}
where $\Big|_{a^{\lambda}}$ means that we expand to power $a^{\lambda}$ and keep only those terms. In ref.~\cite{Bengtsson:1983pd} we give all the generators including the non-linear terms in the other dynamical generators. These expressions are unique if we insist on having a minimal power of the transverse derivatives. We only expect the three-point coupling for the gravity case $\lambda = 2$ to be the beginning of an infinite series of higher-point functions of a self-interacting theory. The higher-spin fields will necessarily couple to each other bringing in all the higher-spins to get a consistent Poincar\'e invariant Hamiltonian.

\section{Counterterms}\label{sec:counterterm}
A necessary condition for the computations above is that the coupling constant $\alpha$ has dimensions ${length}^{(\lambda - 1)}$. This means that from gravity and upwards in spin the coupling constant necessarily has a dimension and that the three-point coupling that we have just shown cannot be unique. More derivatives in the interaction term can be compensated by $\alpha$ to the appropriate power. Let us now specialize to the gravity case and check the next type of terms that can be consistent with Poincar\'e invariance. We now call the coupling constant for its real name $\kappa$ and the fields for $h$ and $\bar h$.

\subsection{One-loop counterterms}
The one-loop three-point counterterms in the Hamiltonian transformation are of order $\kappa^3$ and are quartic in transverse derivatives. A  possible structure of the Hamiltonian transformation that respects helicity is then 
$$ \delta^{\kappa^3}_{ P^-} h~\sim~  \partial\bar\partial^3 h\, h + \partial^3\bar\partial\,h \bar h.$$
As described above we only need to consider the first term $\delta^{\kappa^3}_{P^-} h~\sim~  h\, h$.  Take the Ansatz, (which is not the most general one but enough for our purposes)
\begin{equation}\label{E:hamil-variation}
\delta^{\kappa^3}_{P^-} h~=~ \kappa^{3} \partial^{+n} \left[\,E \partial^{+m}h\, E^{-1} \partial^{+m}h \right]\Big|_{a^{3},\,b} ,
\end{equation}
with the dimension constraint
\begin{equation}\label{E:dimcons}
n\,+\,2\,m~=~3\ .
\end{equation}
The boost transformations at one-loop are of the form
\begin{align}
\delta^{\kappa^3}_{J^-} h &= -\, x\, \delta^{\kappa^3}_{P^-} h + \delta^{\kappa^3}_s h ,&
\delta^{\kappa^3}_{\bar J^-} h &=-\, \bar x\, \delta^{\kappa^3}_{P^-} h + \delta^{\kappa^3}_{\bar s} h .
\end{align}
Thus the boost transformations are determined if we know the spin parts $\delta^{\kappa^3}_{s} h, \delta^{\kappa^3}_{\bar s} h$.
Comparing with the previous calculations, the spin parts of the boosts have to be given by 
\begin{eqnarray}
\delta^{\kappa^3}_s h&=& g_s\,\kappa^{3} \,\partial^{+n} \left[\,E \partial^{+(m-1)}h\, E^{-1} \partial^{+m}h \right]\Big|_{a^{2},\,b},\\
\delta^{\kappa^3}_{\bar s} h&=& g_{\bar s}\,\kappa^{3}\, \partial^{+n} \left[\,E \partial^{+(m-1)}h\, E^{-1} \partial^{+m}h \right]\Big|_{a^{3}},
\end{eqnarray}
where the coefficients $g_s$ and $g_{\bar s}$ will be fixed from commutation relations among dynamical generators.

The ``coherent state-like" formulation makes all the commutations with the kinematical generators straightforward and they do not give any constraints. The non-trivial ones to check are those among the dynamical generators
\begin{align}
[\, \delta_{\bar J^-} \,,\, \delta_{P^-}\,]\, h&=0,\label{dyn1}\\
[\, \delta_{J^-}\,,\, \delta_{P^-} \,] \, h&=0,\label{dyn2}\\
[ \,\delta_{J^-}\,,\, \delta_{\bar J^-} \,]\, h&=0.\label{dyn3}
\end{align}
We perform these calculations in Appendix \ref{B}. The remarkable outcome of these calculations are as follows: \eqref{dyn1} gives 
\begin{equation}\label{oneloopgs}
g_{\bar s}~=~ -2\, (m - 1),
\end{equation}
and \eqref{dyn2} gives
\begin{equation}\label{oneloopgs1}
g_{s}~=~ -2\, (m - 3), 
\end{equation}
while \eqref{dyn3} does not yield any further constraint. The free parameter $m$ is not determined! The technical reason for why this can happen at the one-loop level is that the helicity constraint allows a non-zero term in $\delta^{\kappa^3}_{\bar s} h$, which is not allowed at the tree level. We have a seemingly infinite series of possible three-point couplings that all are consistent with the Poincar\'e invariance to this order in the coupling constant and fields. Many of the terms are related which can be seen by looking at the integral form of them, but still there are far too many.

In the work of 't Hooft and Veltman it was shown that there should exist two different one-loop counterterms. In a covariant Lagrangian description they are proportional to $R^2$ or $R_{\mu \nu} \, R^{\mu \nu}$, since any term proportional to $R_{\mu \nu \rho \sigma} \, R^{\mu \nu \rho \sigma}$can be written in terms of the other two because of the Gauss-Bonnet theorem. It is also clear that both the possible terms would be zero if one uses the mass-shell condition $R_{\mu \nu}=0$.

How do we check if the counterterms that we have constructed would be zero on shell? Here we will use a technique that we introduced when discussing possible counterterms for $5d$ maximally supersymmetric Yang-Mills theory~\cite{Brink:2010ti}. Consider the following relation among coherent-state expressions.
\begin{eqnarray}
\left[\, E \partial^{+m} h \, E^{-1} \partial^{+m} h \,\right] \,\Big|_{a^p, b^q} &=&
\left[\, E \hat{\partial}\hat{\bar\partial}\partial^{+m} h \, E^{-1} \partial^{+m} h 
\,-\,E \hat{\bar\partial}\partial^{+m} h \, E^{-1} \hat{\partial}\partial^{+m} h \right.\\
&&\left.-\,E \hat{\partial}\partial^{+m} h \, E^{-1} \hat{\bar\partial}\partial^{+m} h
\,+\,E \partial^{+m} h \, E^{-1} \hat{\partial}\hat{\bar\partial}\partial^{+m} h
\,\right] \,\Big|_{a^{p-1}, b^{q-1}}.\nonumber
\end{eqnarray}
 Suppose we now consider 
 \be
 \Box \left[\,E\partial^{+m}h\,E^{-1}\partial^{+m}h \right]\,\Big|_{a^p, b^q}= -\frac{1}{2}(\partial^+ \partial^- - \partial \, \bar\partial) \left[\,E\partial^{+m}h\,E^{-1}\partial^{+m}h \right]\,\Big|_{a^p, b^q}
 \ee
and let the derivatives act on the expression. Every time we get $\partial^- h$ we use the equations of motion to write it as $\frac{\partial \, \bar\partial}{\partial^+} h + O(h^2)$.

We then find that
\begin{equation}\label{Box}
\Box \left[\,E\partial^{+m}h\,E^{-1}\partial^{+m}h \right]\,\Big|_{a^p, b^q} ~=~ -\,2\,\left[\,E\partial^{+(m+1)}h\,E^{-1}\partial^{+(m+1)}h \, \right] \,\Big|_{a^{p+1}, b^{q+1}}+ O(h^3)\ .
\end{equation}
Let us use now add in the possible counterterms in the equations of motion to get
\be
i\partial^- h =i \frac{\partial \, \bar\partial}{\partial^+} h -i \kappa \,\partial^{+} \left[\,e^{a \hat{\bar \partial}}  h\, e^{-a \hat{\bar \partial}} h\right] \Big|_{a^{2}}+ f(h, \bar h) + \kappa^{3} \partial^{+n} \left[\,E \partial^{+m}h\, E^{-1} \partial^{+m}h \right]\Big|_{a^{3},\,b}+\ldots
\ee
Let us so use (\ref{Box})  to substitute for the last term to get an equation
\be
-2i\Box h = -i \kappa \,\partial^{+} \left[E h\, E^{-1} h\right] \Big|_{a^{2}}+ f(h, \bar h) - \frac{1}{2}\Box \kappa^{3} \partial^{+n} \left[\,E \partial^{+(m-1)}h\, E^{-1} \partial^{+(m-1)}h \right]\Big|_{a^{2}}+O(h^3).
\ee
We can now make a field redefinition
\be
h' = h + \frac{i}{4}\kappa^{3} \partial^{+n} \left[\,E \partial^{+(m-1)}h\, E^{-1} \partial^{+(m-1)}h \right]\Big|_{a^{2}}\ ,
\ee
to obtain the equation
\be
-2i\Box h' = -i \kappa \,\partial^{+} \left[E h'\, E^{-1} h'\right] \Big|_{a^{2}}+ f(h', \bar h') + O({h'}^3).
\ee
We can now drop the prime and we find only the tree-level equation of motion remaining to this order in the fields. However we still have to tackle the problem with the too many terms, but let us before that look at higher loop orders.

\subsection{Higher-loop 3-point counterterms}
The formalism we have set up makes it completely straightforward to check higher-loop (higher-derivative) counterterms. Let us try the following Ans\"{a}tze for the $l$-loop order 

\begin{eqnarray}
\delta^{\kappa^{2l+1}}_{P^-} h&=&\kappa^{2l+1} \partial^{+n} \left[\,E \partial^{+m}h\, E^{-1} \partial^{+m}h \right]\Big|_{a^{2+l},\,b^l} \ ,\\
\delta^{\kappa^{2l+1}}_s h&=& g_s\,\kappa^{2l+1} \,\partial^{+n} \left[\,E \partial^{+(m-1)}h\, E^{-1} \partial^{+m}h \right]\Big|_{a^{1+l},\,b^l} \ ,\\
\delta^{\kappa^{2l+1}}_{\bar s} h&=& g_{\bar s}\,\kappa^{2l+1}\, \partial^{+n} \left[\,E \partial^{+(m-1)}h\, E^{-1} \partial^{+m}h \right]\Big|_{a^{2+l},\,b^{l-1}} \ ,
\end{eqnarray}
with $n+2m= 2l+1$. Following the calculations performed in the one-loop case in Appendix \ref{B} gives immediately that the only constraints are
\begin{equation}
g_{s}~=~ -2\, (m - l-2)\ , ~\qquad g_{\bar s}~=~ -2\, (m - l).
\end{equation}
As in the one-loop case there is still one parameter $m$ that is free. We can also redo the same calculations as above to show that all of these terms can be absorbed in field redefinitions using the equations of motion.

The key observation that a counterterm will be zero on-shell is that it contains both kinds of transverse derivatives. However, when we get to the two-loop level there is a new type of term that is consistent with the helicity constraint,  namely $\delta^{\kappa^5}_H h~\sim~  {\partial}^6\bar h\, \bar h\ $.

For this Ansatz, one easily sees that there is no possible term with two $\bar h$'s for $\delta^{\kappa^5}_s h$
because of helicity and the numbers of allowed transverse derivatives. The non-linear terms are hence of the form\begin{eqnarray}\label{barhbarhans}
\delta^{\kappa^5}_{P^-} h&=&\kappa^{5} \partial^{+n} \left[\,E \partial^{+m}\bar h\, E^{-1} \partial^{+m}\bar h \right]\Big|_{b^6} \ ,\\
\delta^{\kappa^5}_{\bar s} h&=& g_{\bar s}\,\kappa^{5}\, \partial^{+n} \left[\,E \partial^{+(m-1)}\bar h\, E^{-1} \partial^{+m}\bar h \right]\Big|_{b^5} .
\end{eqnarray}
In Appendix \ref{barhbarh} we go through the relevant commutators with these Ans\"{a}tze. In this case we find the result that there is just one unique solution to \eqref{dyn1}--\eqref{dyn3}.
Indeed, we find 
\begin{align}
m=4, \qquad n=-3, \qquad g_{\bar s} = -4.\nn
\end{align}
Thus we have the solution
\begin{align}~\label{three-loop}
\delta^{\kappa^5}_{P^-} h&=\kappa^{5}\frac{1}{\partial^{+3}} \left[\,E \partial^{+4}\bar h\, E^{-1} \partial^{+4}\bar h \right]\Big|_{b^6} \ ,\\
\delta^{\kappa^5}_{\bar s} h&
= -4\,\kappa^{5}\,\frac{1}{\partial^{+3}} \left[\,E \partial^{+3}\bar h\, E^{-1} \partial^{+4}\bar h \right]\Big|_{b^5} .
\end{align}
From its construction it is clear that this is non-zero on-shell. We do expect to find one such counterterm since the work of Goroff and Sagnotti~\cite{Goroff:1985th}. In a covariant Lagrangian it must correspond to a term proportional to $R_{\mu \nu \rho \sigma}R^{\rho \sigma \tau \eta}R_{ \tau \eta}{}^{\mu \nu}$.
This term \eqref{barhbarhans} can also be extended to any loop order $l$ by changing the $|_{b^6}$ to $|_{a^l, b^{(l+6)}}$ and change the powers of the $\partial^+$'s to get the correct dimension. Also all such counterterms can be shown to be zero on-shell with the same technique as above.

Even before we solve the problem with too many possible counterterm we have found that an on-shell three-point function in gravity has only a singularity at the two-loop level. At any other loop-level the diverging terms will be proportional to a $p^2$ of any of the external legs. This is a fact known in amplitude analysis circles~\cite{Lance, Benincasa:2011pg}.

\section{Contamination from higher-spin fields
and the residual reparametrization invariance}\label{sec:higherspin}
The algebraic approach is approximate until one has computed all higher-order terms. If we had the computational power to check higher-order terms we expect to find that most counterterms we found at the three-point level will not survive. Remember that the transformations are non-linear and when we checked the Poincar\'e algebra we only did it to the three-point level. There are contributions at the four-point level from the counterterms, when we check the algebra,  which should be matched to linear transformations of the four-point counterterms and this goes on ad infinitum. We know the four-point coupling~\cite{Bengtsson:1983vn} which was computed from the covariant formulation. It is quite complicated and unless we can find some better way of summarizing it, the checking of the counterterms at the four-point level is computationally quite hard.

There is a further problem in the formalism. As seen in (\ref{H}) we describe all even-spin fields with a complex index-free field. Consider for example a spin-4 field described covariantly by $\phi^{\mu \nu \rho \sigma}$. In the light-frame formulation we would correspondingly have a field where the indices  run over the transverse components, if we use $SO(2)$ as the helicity symmetry instead of $U(1)$. 
 Such a field has five degrees of freedom with helicities $4, \, 2,\, 0,\,-2,\, -4$. By taking a trace of this field we get the helicity $2$ and $-2$ field and the helicity 0 component. By taking a double trace we get the helicity $0$ field. The helicity $4$ and $-4$ field is obtained by subtracting out the other helicities. This is the normal procedure, but we can also get the helicity $2$ and $-2$ components if we so wish. The field $\phi^{\mu \nu \rho \sigma}$ has naturally a three-point self-interaction with four derivatives. Neither the helicities $4$ and $-4$ nor the helicities $2$ and $-2$ will be a consistent self-interacting field theory to all orders. (Note that in our formalism where we use $U(1)$ as the helicity symmetry an index free field $\phi$ can describe any helicity $\lambda$ and $-\lambda$.) However, they can both have a consistent three-point coupling to this order. The one for the helicities $2$ and $-2$ must then be among the counterterms we found in (\ref{E:hamil-variation}). 

It is now meaningless to try to explain all the unwanted counterterms in (\ref{E:hamil-variation}). The question is if we have missed any symmetry that can be used to give a unique answer. Indeed we have. Even if we believe that the algebraic approach will give the unique theory given the computational needs to compute to all orders, there is indeed a further symmetry in the $lc_2$ formulation, the remaining gauge invariance. This is a hint that there is more to gravity than meets the eye. Note that the light-cone gauge choice is a physical one and hence there is no BRST symmetry in this gauge. Let us now follow the steps from a covariant formulation to see what happens to the reparametrization invariance.

Recall that in the covariant formulation we write the metric tensor in terms of the gravity field as 
\begin{align}\label{gtoh}
g_{\mu\nu} = \eta_{\mu\nu} + \kappa\, h_{\mu\nu} ,
\end{align}
where $h_{\mu\nu}$ is symmetric. Its inverse is defined so that $ g^{\mu\rho} g_{\rho\nu} = \delta^{\mu}{}_{\nu}$ and thus yielding
\begin{align}\label{inverse}
g^{\mu\nu}=\eta^{\mu\nu} -\kappa h^{\mu\nu} + \kappa^2 h^{\mu\rho}h_{\rho}{}^\nu - \kappa^3 h^{\mu\rho}h_{\rho}{}^{\sigma}h_{\sigma}{}^\nu +\cdots,
\end{align}
where the indices for $h_{\mu\nu}$ are raised and lowered through $\eta^{\mu\nu}=\mathrm{diag}(-,+,+,+)$, e.g., $h^{\mu\nu}=\eta^{\mu\rho}\eta^{\nu\sigma}h_{\rho\sigma}$.

One can fix the gauge as $g_{--}=g_{-i}=0, ~g_{-+}=-1$, or equivalently  
\begin{align}\label{lcgauge}
h^{+\mu}=0.
\end{align}
This eliminates four degrees of freedom.
To eliminate all the unphysical degrees of freedom, we use equations of motion to eliminate these degrees of freedom in terms of the physical ones. This procedure will generate higher order interactions that are not in the covariant theory. The equation of motion that we use is Einstein equation for vacuum, $R_{\mu\nu}=0$ . It follows from
\begin{align}
R_{--}=0&\quad \Longrightarrow\qquad  {\partial^{+2}}h^i{}_i =0\quad {\rm ( to~ lowest~ order ~in ~\kappa)}
\end{align}
that   $h_{ij}$ is traceless ($h^i{}_i=0$). Once the traceless condition is implemented, the Einstein equation for each component can be summarized as, to lowest order, 
\begin{align}\label{eom}
R_{-i}= 0 &\quad\to\quad h^{-i} = \frac{\partial_j}{\partial^+}h^{ij},\cr
R_{+-}= 0 &\quad\to\quad h^{--} = \frac{1}{\partial^{+2}}\partial_i\partial_jh^{ij},\cr
R_{+j}=0 &\quad\to\quad \frac{1}{\partial^{+2}}\partial_i\Box h^{ij}=0,\cr
R_{ij}=0&\quad\to\quad-\frac{\kappa}{2}\Box h_{ij} = 0,\cr
R_{++}=0&\quad\to\quad-\frac{\kappa}{2}\frac{\partial_i\partial_j}{\partial^{+2}}\Box h^{ij} = 0,
\end{align}
which agree with \cite{Bengtsson:1983pe}. The first two equations show that the unphysical degrees of freedom, $h^{-i}$ and $h^{--}$, can be replaced by the physical ones $h^{ij}$, and the other equations are proportional to the equation of motion for the physical degrees of freedom, $\Box h^{ij}$. The higher-order terms will lead to new interaction terms in the physical fields.

This leaves us with only two degrees of freedom, say, $h_{11}$ and $h_{12}$. One then defines
\begin{align}
	h &= \frac{1}{\sqrt2} ( h_{11} + i h_{12} ), &
 \bar h &= \frac{1}{\sqrt2} ( h_{11} - i h_{12} ),
\end{align}
or
\begin{align}
h_{11} &= \frac{1}{\sqrt2} (h + \bar h), &
h_{12} &= -\frac{i}{\sqrt2}(h - \bar h),
\end{align}
and the dynamical equations will now only contain $h$ and $\bar h$. 

We can now ask what happens to the reparametrizations with a parameter $\xi^{\mu}(x^+,x^-, x, \bar x)$. From the gauge fixings it is straightforward that the parameter for the residual reparametrizations must satisfy $\partial^+ \xi^\mu = 0$. It is also clear that since we only have $h_{11}$ and $h_{12}$ we are only interested in transformations with the parameters $\xi^{1,2}$. As usual we combine them to $\xi$  and $\bar \xi$. Finally we have to check what constraints the equations that eliminate the unphysical degrees of freedom give us. We find that a remaining infinitesimal reparametrization invariance is the following,
\be \label{full symmetry}
\delta h = \frac{1}{2\kappa} \partial \xi + \xi  \bar \partial h + \bar \xi \partial h,
\ee
where $\xi$ satisfies 
\be
\bar \partial  \xi = 0,
\ee
i.e. $\xi = \xi (\bar x)$.
The transformation of $\bar h$ is obtained by complex conjugation. We see that this is a two-dimensional reparametrization $x\rightarrow x+\xi (\bar x)$ and $ \bar x \rightarrow \bar x + \bar \xi( x)$. These look like two-dimensional conformal transformations but with $x$ and $\bar x$ interchanged in the transformations. Alas, this is not a transformation that can be closed to generate a finite symmetry. This can also be seen by commuting two transformations (\ref{full symmetry}). Formally it looks as if it closes with a parameter $\xi_{12} =\frac{1}{2\kappa}( \bar \xi_2 \partial \xi_1 - \bar \xi_1 \partial \xi_2)$. However, it is obvious that this expression satisfies $\bar\partial  \xi_{12} \neq 0$. Can the transformations (\ref{full symmetry}) still be used? The answer is yes.
 
 Consider the action after the gauge choice $h^{+ \mu} = 0$ is implemented. It can be written as  
 \be \label{action}
 S = \int d^4x\,\, ( {\cal L}( h, \bar h) + {\cal L}_{alg}).
 \ee
In order to get to this form, we have put the expression in a form such that the unphysical fields can be integrated out.  Note that we only expand the action up to three-point functions. The kinetic terms of the unphysical fields are quadratic expressions in them which do not involve any time-derivative $\partial^-$. The expressions with their three-point couplings can be completed to squares in order to make the functional integral simple. ${\cal L}_{alg}$ then consists of the unphysical fields also coupled to the physical one.  As can be seen from the first two equations of \eqref{eom} the functional integration will only contribute a determinant in $\partial^+$   to the order that we consider and can hence be disregarded. However, for the arguments that will follow we keep ${\cal L}_{alg}$. For a detailed analysis of the corresponding action in $N=4$ Yang-Mills theory, see~\cite{Brink:1982pd}. 
By taking functional derivatives of ${\cal L}( h, \bar h) $ we will get the equations of motion and hence the expressions $\delta _{P^-} h$ from above. Both terms of the action are invariant separately under the infinitesimal transformations (\ref{full symmetry}). It is only if we want to look for the finite symmetry in (\ref{action}) that we will be led to reparametrizations with parameters $\xi^\mu$ satisfying $\partial^+\xi^\mu=0$,  and that is the true symmetry of the problem.
 
 We are used to the fact that open algebras usually close on the equations of motion. If we start with the transformations (\ref{full symmetry}) and try to close it, we will in the end be led to the bigger symmetry and to prove its closure we will have to use the equations of motion for the unphysical fields as they can be derived from ${\cal L}_{alg}$ in the action above. In this sense the transformations close on-shell. 
 
 Consider now adding counterterms to (\ref{action}). They will be generated from the dynamical part which is ${\cal L}( h, \bar h)$ and the algebraic part ${\cal L}_{alg}$ will not be affected, since in a perturbative loop calculation only the fields $h$ and $\bar h$ take part. In the new action
 
 \be \label{ct}
 S = \int d^4x\,\, ( {\cal L}( h, \bar h) + {\cal L}^{ct}( h, \bar h) +{\cal L}_{alg}).
 \ee
${\cal L}_{alg}$ is still invariant under the infinitesimal transformation (\ref{full symmetry}). Hence the first two terms ${\cal L}( h, \bar h) + {\cal L}^{ct}( h, \bar h)$ must also  be invariant under those transformations. This is the reason why we can use (\ref{full symmetry}) to distinguish between the counterterms coming from spin-$2$ and the ones coming from other spins.

 \subsection{Remaining reparametrization invariance of the counterterms}
 
 In order to select the correct counterterms we have to check which terms do indeed satisfy the remaining reparametrization invariance. Since the infinitesimal transformations (\ref{full symmetry}) are non-linear they connect terms with different number of fields $h$. The first counterterm has no lower-order term to talk to since the transformations connect terms with the same number of derivatives. Hence it must be annihilated by the inhomogeneous term $\delta h  = \frac{1}{2\kappa} \partial \xi$.
 
 We start with the one-loop terms \eqref{E:hamil-variation}
\begin{align}
\delta H^{\kappa^3} &= \kappa^3 
\int d^4x~\delta \left(\partial^+\bar h\, \partial^{+(3-2m)} \left[\,E \partial^{+m}h\, E^{-1} \partial^{+m}h \right]\Big|_{a^{3},\,b} 
\right).
\end{align}
Consider first the case $m=0$. The expression is then
\be
\delta H^{\kappa^3}= -\kappa^3 
\int d^4x~\delta \left( \partial^{+4} \bar h\,   \left[\,E  h\, E^{-1}  h \right]\Big|_{a^{3},\,b}\right) .
\ee
 We see that by varying $\delta \bar h=\frac{1}{2\kappa} \bar \partial \bar \xi$ that this term will be zero since $\partial^+ \bar \xi =0$. The term to worry about inside the coherent state-like expression is when one gets terms with $\partial \xi$, that is not multiplied by $\bar \partial$ or/and $\partial^+$. In the expansions of the coherent state-like expression we see that such terms occur and this term is not invariant. We easily see that we need $m \geq 2$. Consider so the case $m=2$. In this case the term to worry about is the variation of $\bar h$. It is now not multiplied by a $\partial^+$. We are hence left with a term
\be
\delta H^{\kappa^3}= -\frac{1}{2} \kappa^2 
\int d^4x~ \bar \xi\,   \left[\,E \partial^{+2} h\, E^{-1} \partial^{+2} h \right]\Big|_{a^{3},\,b} .
\ee
Here both $\partial$ and $\partial^+$ can be partially integrated and we see that the terms we get by expanding the expression inside the bracket can be partially integrated to cancel each other pairwise. In the case $m>2$ we cannot partially integrate $\partial^+$ and the expression is easily seen to be non-zero. We have hence seen that there is only one one-loop counterterm that is consistent with the residual reparametrization invariance, namely
\be
H^{\kappa^3} = -\kappa^3 
\int d^4x~   \bar h\,   \left[\,E \partial^{+2} h\, E^{-1} \partial^{+2} h \right]\Big|_{a^{3},\,b}, 
\ee
or in the ``equation of motion" from~(\ref{E:hamil-variation})
\begin{equation} 
\delta^{\kappa^3}_{P^-} h~=~ \kappa^{3} \frac1{\partial^+} \left[\,E \partial^{+2}h\, E^{-1} \partial^{+2}h \right]\Big|_{a^{3},\,b} .
\end{equation}
The fact that only one term survives might look a bit puzzling at the first moment. We know and have argued earlier in the paper that there should be two terms. However if we consider the two terms  $R^2$ and $R_{\mu \nu} \, R^{\mu \nu}$ and reduce them to the $lc_2$ formulation we will find that they contain the same three-point coupling. They only differ in higher-order terms. Hence we have understood the counterterms at the one-loop level.

 At the two-loop level we only found one possible term with six transverse derivatives which is non-zero on-shell. This term must then satisfy the correct remaining reparametrization. Here we give explicitly that calculation. The variation on two-loop counterterm that we found (\ref{three-loop}) is  
\begin{align}\label{2loopinv}
\delta H^{\kappa^5}&= \kappa^5 \int d^4x~\frac{1}{\partial^{+2}}\delta \bar{h}\Big[E\partial^{+4}\bar{h}E^{-1}\partial^{+4}\bar{h}\Big]_{b^6}\\
&\sim\kappa^4\int d^4x~\frac{1}{\partial^{+2}}\bar \partial \bar\xi
\Big(2\partial^{+4}\bar{h}\frac{\partial^6}{\partial^{+2}}\bar{h}
-12\partial^{+3}\partial\bar{h}\frac{\partial^5}{\partial^{+}}\bar{h}
+30\partial^{+2}\partial^2\bar{h}\partial^4\bar{h}
-20\partial^{+}\partial^3\bar{h}\partial^{+}\partial^3\bar{h}\Big).\nn
\end{align}
The invariance of this counterterm under the inhomogenous part of the variation of the remaining gauge transformation can be seen from repeated use of partial integration with respect to $1/\partial^+$. Cancellation of terms follows from that the coefficients of each term above come binomial expansion. An elegant way to see this is to use the identities of the coherent-like forms (See Appendix \ref{appA}). Using the identity \eqref{id4-2}
\begin{align}
E^{-1} \Big[  E f_1 \, f_2\Big]\,\Big|_{b^6}~=~\frac{1}{\partial^{+p}} \Big[ E f_1\, E^{-1} \partial^{+p} f_2 \Big] \,\Big|_{b^6}, 
\end{align}
one can re-express \eqref{2loopinv} as
\begin{align}
\kappa^5\int d^4x\frac{1}{\partial^{+2}}\bar\partial\bar{\xi}E^{-1}\partial^{+6}\Big[E\partial^{+4}\bar{h}\frac{1}{\partial^{+2}}\bar{h}\Big]_{b^6}.
\end{align}
After the integrations by parts, it becomes
\begin{align}
\kappa^5\int d^4xE^{-1}\partial^{+6} \Big(\frac{1}{\partial^{+2}}\bar\partial\bar{\xi}\Big)\Big[E\partial^{+4}\bar{h}\frac{1}{\partial^{+2}}\bar{h}\Big]_{b^6},
\end{align}
which vanishes because of the gauge constraint $\partial^+\bar\xi =0=\partial\bar\partial\bar\xi$.

\section{Some Consequences from the Virasoro-like Symmetry.}\label{sec:virasoro}

In the previous sections we have argued that the transformations (\ref{full symmetry}) is important in order to find the form for the counterterms. We will here show that it has further consequences also for the form of the Hamiltonian. Introduce the following ``covariant derivatives"
\begin{align}\label{eq:CovD}
{\cal D}\bar h=\partial\bar h+\frac{2\kappa}{\partial^{+2}}\Big[h\bar\partial\partial^{+2}\bar h-\frac{\bar\partial}{\partial^+}h\partial^{+3}\bar h\Big]+O(h^3),\\
{\cal\bar D}h=\bar\partial h+\frac{2\kappa}{\partial^{+2}}\Big[\bar h\partial\partial^{+2}h-\frac{\partial}{\partial^+}\bar h\partial^{+3} h\Big]+O(h^3).
\end{align}
We can then write the Hamiltonian at least up the to three-point level as 
\be
H=P^- = \int d^4x {\cal D}\bar h {\cal\bar D}h.
\ee
We expect it to be true to all orders.  We can check that ${\cal\bar D}h$ transforms covariantly under (\ref{full symmetry}). Since the transformation parameter $\xi$ satisfies $\partial^+ \xi =0$ we can easily see that the transformations do not uniquely determine the form of the ``covariant derivatives", since different powers of $\partial^+$ properly distributed can give terms that transform the same. To select between them we have to check the transformations under the Poincar\'e transformations.

We could use this form in the functional integral and derive expressions for the S-matrix to see if they have a better UV-property than what meets the eye from power counting. We might come back to it but since we know that gravity is diverging in the UV we will instead use our knowledge and take it over to the $N=8$ Supergravity theory.

\section{Extension to $N=8$ Supergravity}\label{sec:n8sugra}
It would take us too far into detail to also do the same programme for the $N=8$ Supergravity theory in this paper. Here we will just sketch how this programme will work and be content to show that there are no three-point counterterms in this theory,  a fact which has been known since the early days of supergravity~\cite{Deser:1977nt}. The detailed work to look for higher-point counterterms will be delayed to future papers. We begin by giving a brief overview of the light-frame formulation of the  $N=8$ theory~\cite{Bengtsson:1983pg, Brink:2008qc}.

To formulate ${N}=8$ Supergravity on the light-cone, 
one considers the superspace spanned by eight Grassmann variables, $\theta^m$ and their complex conjugates $\bar\theta_m$ $(m=1,...,8)$, on which $SU(8)$ acts linearly. 
The chiral derivatives are defined as 
\bea
d^m ~\equiv~ -\frac{\p}{\p\bar\theta_m} -\frac{i}{\sqrt{2}} \theta^m\p^+ \, , ~~~ 
\bar d_m ~\equiv~ \frac{\p}{\p\theta^m} +\frac{i}{\sqrt{2}} \bar\theta_m\p^+ \ , 
\eea
which satisfy canonical anticommutation relations

\begin{equation}\label{phianticomm}
\left\{ d^m\,,\,\bar d_n\right\}~=~ -i \sqrt{2}\delta^m{}_{n} \p^+ \ .
\end{equation}
The $256$ physical degrees of freedom of ${N}=8$ Supergravity are 
the spin-2 graviton $h$ and $\overline h$,  eight spin-$\frac{3}{2}$ gravitinos ${\psi}^m$ and 
${\overline \psi}_m$, twenty eight vector fields 
\begin{align}
 \overline B_{mn} \equiv \frac{1}{\sqrt{2}} \left( B^1_{mn} \,+\,i\,B^2_{mn}\right),
\end{align}
and their conjugates, fifty six  gauginos ${\overline \chi}_{mnp}$ and $\chi^{mnp}$, and finally seventy real scalars 
\begin{align}\label{Dsd}
{\overline D}_{mnpq} =\frac{1}{4!}\epsilon_{mnpqabcd}D^{abcd}.
\end{align} 
They are all contained in {\em one} constrained chiral superfield 
\begin{align}
\phi(y)=&~\frac{1}{{\p^+}^2}\,h\,(y)\,+\,i\,\theta^m\,\frac{1}{{\p^+}^2}\,{\overline \psi}_m\,(y)\,+\,i\,\theta^{mn}_{}\,\frac{1}{\p^+}\,{\overline B}_{mn}\,(y)\cr
\;&-\,\theta^{mnp}_{}\,\frac{1}{\p^+}\,{\overline \chi}^{}_{mnp}\,(y)\,-\,\theta^{mnpq}_{}\,{\overline D}^{}_{mnpq}\,(y)+\,i\widetilde\theta^{}_{~mnp}\,\chi^{mnp}\,(y)\cr
&+\,i\widetilde\theta^{}_{~mn}\,\p^+\,B^{mn}\,(y)+\,\widetilde\theta^{}_{~m}\,\p^+\,\psi^m_{}\,(y)+\,{4}\,\widetilde\theta\,{\p^+}^2\,{\bar h}\,(y) ,
\end{align}
where the bar denotes complex conjugation, and 
\begin{align}
\theta^{a_1a_2...a_n}_{}~=~\frac{1}{n!}\,\theta^{a_1}\theta^{a_2}_{}\cdots\theta^{a_n}_{}\ ,\quad \widetilde\theta^{}_{~a_1a_2...a_{n}}~=~ \epsilon^{}_{a_1a_2...a_{n}b_1b_2...b_{(8-n)}}\,\theta_{}^{b_1b_2\cdots b_{(8-n)}}.
\end{align}
Here one uses the chiral coordinates 
\begin{align}
y~=~(x,\,\bar x,\, x^+, \,y^-\equiv x^- -\frac{i}{\sqrt2}\theta^m\bar\theta_m\, )
\end{align}
so that  $\phi$ and its complex conjugate $\overline \phi$ satisfy the chiral constraints
\bea\label{chiralconstraints}
d_{}^m\, \phi ~=~0, \qquad \overline d^{}_m\, \overline \phi ~=~0.
\eea
The complex chiral superfield is related to its complex conjugate by the {\em inside-out constraint}
\begin{equation}\label{insideout}
\phi~=~\frac{1}{4\,\p^{+4}}\,d_{}^1d_{}^2\cdots d_{}^8\, \overline\phi,
\end{equation}
in accordance with the duality condition of ${ D}^{mnpq}$ \eqref{Dsd}. 

It is now straightforward to see that there are no three-point counterterms in this theory. Remember that the two-loop counterterm in gravity is of the form
\be
\delta_{P_-} h \sim \bar h \,\, \bar h.
\ee
If there were a similar counterterm in $N=8$ Supergravity it must be of the form
\be
\delta_{P_-} \phi \sim \bar\phi \,\, \bar\phi.
\ee
However such an expression does not satisfy the chirality constraint (\ref{chiralconstraints}).

As shown in \cite{Brink:2008qc} the integrated Hamiltonian can be written as a quadratic form
\be
H=P^-= \int d^4x \, d^8 \theta \, d^8 \bar \theta \,\, \overline{\delta_{{q_-}} \phi}\, \frac{1}{{\partial^+}^3}\, \delta_{{q_-}} \phi,
\ee
where $\delta_{{q_-}} \phi$ is  the dynamical supersymmetry transformation of  the superfield $\phi$. This is one of the non-linear transformations of the superPoincar\'e algebra. Besides that, it commutes with the $E_7$ transformations~\cite{Brink:2008qc}.
We can now look for counterterms in the transformation $\delta_{{q_-}} \phi$ that not only transform correctly under the full superPoincar\'e algebra but also commute with the $E_7$ transformations and satisfy the conditions given by the remaining gauge invariance. Since it is only the non-linear term that is important for the lowest lying counterterm it means that we should demand that it is annihilated by
\be
\delta h = \partial \xi, 
\ee
but also 
\be
\delta {\overline \psi}_m = \partial \bar \epsilon_m
\ee
and 
\be
\delta {\overline B}_{mn}=\partial \bar \Lambda_{mn}.
\ee

These conditions can be cast into a superspace language and the whole procedure can be done in superspace. We will come back to a more detailed study in the future.
\section{Conclusion}
In this paper we have studied possible counterterms for ordinary gravity in the light-front formulation, the $\it lc_2$ formulation (the light-cone gauge formulation), which only contains the physical degrees of freedom, the helicity $2$ field $h(x)$ and the helicity $-2$ field $\bar h(x)$. This formalism is a non-linear representation of the Poincar\'e  algebra and all the generators that take us forward in the time $x^+$ get non-linear contributions. We find that by making general Ans\"atze for the non-linear terms that the closure of the algebra is not enough to select the possible counterterms. We understand this as a ``contamination" of higher-spin fields with helicities $2$ and $-2$ that can show up at this stage of the calculation. We hence need some kind of symmetry to select the correct gravity terms. In principle the $\it lc_2$ formulation is a fully gauge fixed one. However, we can find an infinitesimal local transformation in the transverse plane under which the Hamiltonian is invariant. We can then understand that this transformation is part of the remaining reparametrization invariance, which exist after the gauge condition is chosen. In order to see the full invariance one has to add back all the unphysical degrees of freedom to the action. However, we can argue that it is enough to check the infinitesimal transformations since the addition of the counterterms does not change the part of the action which is quadratic in the unphysical fields. Hence also the counterterms should be invariant under the infinitesimal local transformations.

Perturbative quantum gravity in the Minkowski space clearly demands more than the classical theory. Computing a certain S-matrix element to a certain loop-order will only include terms up to a certain order in the expansion. Our formulation is unitary, causal and Poincar\'e invariant. By just using these criteria we do not get a unique result. This shows that gravity in Minkowski space contains more than just a unitary, causal and Poincar\'e invariant theory of spin-$2$ particles. This is clearly important for the quantum properties of the S-matrix.

We have also shown how to take this formalism to the $N=8$ Supergravity theory and we have sketched on how to check for possible counterterms in that theory. We intend to come back to that issue in the future.

\acknowledgments
LB would like to thank Edward Witten and Peter Goddard for an invitation to the Institute of Advanced Study where part of this work was done, as well as Slava Mukhanov and Dieter L\"ust for an invitation to the Ludwig-Maximilian-Universit\"at in Munich where another part of the work was done.
SSK thanks Chalmers University of Technology where this work was initiated and he also thanks Universit\'{e} Libre de Bruxelles where part of work was done. 
The authors thank Pierre Ramond, Hermann Nicolai, Chris Hull and Marc Henneaux for useful discussions.

\appendix
\section{Useful Identities}\label{appA}

We introduced the ``coherent state-like" form of the Ans\"atze for the dynamical generators in Section \ref{sec:LCformulations}.
We here write some important identities which are useful for computing the commutation relations among the dynamical generators.\\

\noindent$\bullet$ Identity 1
\begin{align}
\sum^n_{r=0}\,(-1)^r\,
\binom{n}{r}
 x^{n-r}h \, x^r h~=~\left(\frac{\partial}{\partial a}\right)^n \left[ e^{ax}h\,  e^{-ax}h\right]\Bigg|_{a=0} .
\end{align}
One can take $\hat \partial$ or $\hat{\bar\partial}$ instead of $x$. For instance, 
\begin{align}
\sum^\lambda_{n=0}\,(-1)^n\,\binom{\lambda}{n}\left[ \,
{\hat{\bar\partial}}^{(\lambda-n)}h\, {\hat{\bar\partial}}^{n}h\, \right]~=~\left(\frac{\partial}{\partial a}\right)^\lambda
 \left[ \,e^{a\hat{\bar\partial}}h\,  e^{-a\hat{\bar\partial}}h\,\right]\Bigg|_{a=0}.
\end{align}
where 
\begin{align}
\hat{\bar \partial}\equiv\frac{ \bar \partial}{\partial^+}\qquad {\rm and}\qquad \hat{\partial}\equiv\frac{\partial}{\partial^+}.
\end{align}
These expressions frequently appear in \cite{Bengtsson:1983pd} with a helicity $\lambda$ field $h$.
For convenience, we now define 
\begin{align}
 \left[ e^{ax}h\,  e^{-ax}h\right]\Bigg|_{a^n}~:=~\left(\frac{\partial}{\partial a}\right)^n \left[ e^{ax}h\,  e^{-ax}h\right]\Bigg|_{a=0}.
\end{align}\\

\noindent$\bullet$ Identity 2
\begin{align}
\left(\frac{\partial}{\partial a}\right)^n \left[\, a \,f(a)\,\right]\, \Bigg|_{a=0} ~=~ n \left(\frac{\partial}{\partial a}\right)^{n-1} \left[\,f(a)\,\right]\,\Bigg|_{a=0},
\end{align}
or
\begin{align} \label{A6}
\left[\, a \,f(a)\,\right]\, \Big|_{a^n}~=~n \left[\,f(a)\,\right]\,\Big|_{a^{n-1}}.
\end{align}
\\
As a corollary, one finds that if
\begin{align}
\delta^g_{P^-} h ~=~ \partial^{+n}\left[ E \partial^{+m} h \, E^{-1} \partial^{+m} h \right] \,\Big|_{a^p, b^q},
\end{align}
where  
\begin{align}
E:= \exp( a \hat{\bar\partial} + b\hat{\partial}), &&E^{-1}:= \exp( -a \hat{\bar\partial} - b\hat{\partial})
\end{align}
and $p+q$ is even,
then 
\begin{align}
\partial^{+n}\left[2 a E \hat{\bar\partial}^2\partial^{+m} h \, E^{-1} \partial^{+m} h \right] \,\Big|_{a^p, b^q}
~=~p \,\delta_{\hat{\bar\partial}}\delta^g_{P^-} h.
\end{align}
\underline{Proof:}\\
Consider
\begin{align}
\delta_{\hat{\bar\partial}}\delta^g_{P^-} h&= \left(\frac{\partial}{\partial b}\right)^{q}\left(\frac{\partial}{\partial a}\right)^{p} \partial^{+n}\left[2 E \hat{\bar\partial}\partial^{+m} h \, E^{-1} \partial^{+m} h \right] \,\Big|_{a=0=b}\cr
&= \left(\frac{\partial}{\partial b}\right)^{q}\left(\frac{\partial}{\partial a}\right)^{p-1} \partial^{+n}\left[2 E \hat{\bar\partial}^2\partial^{+m} h \, E^{-1} \partial^{+m} h \right] \,\Big|_{a=0=b}\cr
&= \partial^{+n} \left[2 E \hat{\bar\partial}^2\partial^{+m} h \, E^{-1} \partial^{+m} h \right] \,\Big|_{a^{p-1}, b^q},
\end{align}
where, in the second line, we have used 
\begin{align}
\left(\frac{\partial}{\partial b}\right)^{q}\left(\frac{\partial}{\partial a}\right)^{p-1} \partial^{+n}\left[2 E \hat{\bar\partial}\partial^{+m} h \, E^{-1} \hat{\bar\partial}\partial^{+m} h \right] \,\Big|_{a=0=b} ~=~0,
\end{align}
due to symmetry that each term is the same form ($\hat{\bar\partial}\partial^{+m} h$) but one restricts oneself to an odd ($p+q-1$) powers. It is then easy to see   
\begin{align}
\partial^{+n}\left[2 aE \hat{\bar\partial}^2\partial^{+m} h \, E^{-1} \partial^{+m} h \right] \,\Big|_{a^p, b^q}~=~
p\, \partial^{+n} \left[2 E \hat{\bar\partial}^2\partial^{+m} h \, E^{-1} \partial^{+m} h \right] \,\Big|_{a^{p-1}, b^q}~=~p\, \delta_{\hat{\bar\partial}}\delta^g_H h.
\end{align}
In general, for $p+q$ odd 
\begin{align}
\delta^g_s h ~=~ \partial^{+n}\left[ E \partial^{+m} h \, E^{-1} \partial^{+m'} h \right] \,\Big|_{a^p, b^q},
\end{align}
one then finds that 
\begin{align}
\partial^{+n}\left[ a E \hat{\bar\partial}^2\partial^{+m} h \, E^{-1} \partial^{+m'} h
-a E \partial^{+m} h \, E^{-1}\hat{\bar\partial}^2 \partial^{+m'} h
 \right] \,\Big|_{a^p, b^q}
=p \,\delta_{\hat{\bar\partial}}\delta^g_s h,
\end{align} 
and
\begin{align}\label{A14}
\partial^{+n}\left[ b E \hat{\partial}^2\partial^{+m} h \, E^{-1} \partial^{+m'} h
-b E \partial^{+m} h \, E^{-1}\hat{\partial}^2 \partial^{+m'} h
 \right] \,\Big|_{a^p, b^q}
=q \,\delta_{\hat{\bar\partial}}\delta^g_s h.
\end{align} \\

\noindent$\bullet$ Identity 3\\
From expansion of the exponentials, it is clear that 
\begin{align}
\label{a8}
\left[\, E \partial^{+m} h \, E^{-1} \partial^{+m} h \,\right] \,\Big|_{a^p, b^q} =&
\left[\, E \hat{\partial}\hat{\bar\partial}\partial^{+m} h \, E^{-1} \partial^{+m} h 
\,-\,E \hat{\bar\partial}\partial^{+m} h \, E^{-1} \hat{\partial}\partial^{+m} h \right.\\
&\left.-\,E \hat{\partial}\partial^{+m} h \, E^{-1} \hat{\bar\partial}\partial^{+m} h
\,+\,E \partial^{+m} h \, E^{-1} \hat{\partial}\hat{\bar\partial}\partial^{+m} h
\,\right] \,\Big|_{a^{p-1}, b^{q-1}}.\nonumber
\end{align}
For $p+q$ even, \eqref{a8} simplifies to
\begin{equation}
\left[\, E \partial^{+m} h \, E^{-1} \partial^{+m} h \,\right] \,\Big|_{a^p, b^q} ~=~2\,
\left[\, E \hat{\partial}\hat{\bar\partial}\partial^{+m} h \, E^{-1} \partial^{+m} h 
\,-\,E \hat{\bar\partial}\partial^{+m} h \, E^{-1} \hat{\partial}\partial^{+m} h 
\,\right] \,\Big|_{a^{p-1}, b^{q-1}} \ .
\end{equation}
As a corollary, 
\begin{equation}
\Box \left[\,E\partial^{+m}h\,E^{-1}\partial^{+m}h \right]\,\Big|_{a^p, b^q} ~=~ -\,2\,\left[\,E\partial^{+(m+1)}h\,E^{-1}\partial^{+(m+1)}h \, \right] \,\Big|_{a^{p+1}, b^{q+1}}  +\mathcal{O}(h^3)\ .
\end{equation}\\

\noindent$\bullet$ Identity 4\\
A relevant identity from the binomial expansion \cite{Bengtsson:1983pd} is  
\begin{align} \label{id4}
&\sum^{\lambda}_{n=0} (-1)^n \binom{\lambda}{n}\frac{\partial^n}{\partial^{+(n+1)}}\left( \frac{\partial^{\lambda-n}}{\partial^{+(\lambda-n)}}\bar h\,\partial^{+\lambda}h\right)
=\frac{1}{\partial^{+(\lambda+1)}}\sum^{\lambda}_{n=0} \,(-1)^n
\binom{\lambda}{n}\left( \frac{\partial^{\lambda-n}}{\partial^{+(\lambda-n)}}\bar h\,\partial^{+(2\lambda-n)}\partial^n h\right).
\end{align}
In terms of the coherent state-like form, this identity is expressed as
\begin{eqnarray}\label{id4-1}
E^{-1} \left[  E \bar h \, \partial^{+\lambda} h\right]\,\Big|_{\hat{\partial}^{\lambda}}~=~\frac{1}{\partial^{+\lambda}} \left[ E \bar h \, E^{-1} \partial^{+2\lambda} h \right] \,\Big|_{\hat{\partial}^{\lambda}} \ .
\end{eqnarray}
More generally, 
\begin{eqnarray}\label{id4-2}
E^{-1} \left[  E \bar h \, h\right]\,\Big|_{\hat{\partial}^{p}}~=~\frac{1}{\partial^{+p}} \left[ E \bar h \, E^{-1} \partial^{+p} h \right] \,\Big|_{\hat{\partial}^{p}} \ .
\end{eqnarray}\\

\section{Detailed calculations: one-loop counterterms}\label{B}

In this Appendix we show the detailed computations for the one-loop counterterms in Section \ref{sec:counterterm}.
Here we compute the commutation relations among the dynamical generators $ \bar J^-, J^-$ and ${P}^-$. It follows from the Poincar\'e symmetry algebra that these commutation relations vanish (to all orders).  
Before we compute, we set up the notations. The dynamical generator transformations $\m G$ that act directly on the field are expressed as
\begin{align}
\delta_{\m G^-}h =\delta^{\rm free}_{\m G^-}h  + \delta^{g}_{\m G^-}h.
\end{align}
The free parts of the dynamical generators are
\begin{align}
\delta^{\rm free}_{P^-}h &= -i\frac{\partial\bar\partial}{\partial^+}h,\cr
\delta^{\rm free}_{J^-} h &= i\left(x\,\frac{\partial\bar\partial}{\partial^+} -x^- \partial - \lambda \frac{\partial}{\partial^+} \right) h,\cr 
\delta^{\rm free}_{\bar J^-} h &= i\left( \bar x\, \frac{\partial\bar\partial}{\partial^+}  -x^- \bar \partial + \lambda \frac{\bar \partial}{\partial^+} \right) h,
\end{align}
where $\lambda$ denotes the helicity of the field that the generators act on. For example, $\lambda=2$ for $h$ while $\lambda=-2$ for $h$. 
The Ansatz for the counterterm to power $\kappa^3$  of the dynamical generator $P^-$ \eqref{E:hamil-variation}
\begin{align}\label{1loopans}
\delta^{\kappa^3}_{P^-} h&=~ \kappa^{3} \partial^{+n} \left[\,E \partial^{+m}h\, E^{-1} \partial^{+m}h \right]\Big|_{a^{3},\,b} ,
\end{align}
which is subject to 
\begin{align}\label{dimconst}
n+2m = 3,
\end{align}
which we called the dimension constraint \eqref{E:dimcons}. Other Ans\"atze for the dynamical generators are
\begin{align}\label{1ansother}
\delta^{\kappa^3}_{\bar J^-} h &= -x \delta^{\kappa^3}_{P^-}h + \delta^{\kappa^3}_{s}h; 
\qquad
\delta^{\kappa^3}_{s} h= g_{s}\,\kappa^{3}\, \partial^{+n} \left[\,E \partial^{+(m-1)}h\, E^{-1} \partial^{+m}h \right]\Big|_{a^{2},\, b},\cr
\delta^{\kappa^3}_{\bar J^-} h &= -\bar x \delta^{\kappa^3}_{P^-}h + \delta^{\kappa^3}_{\bar s}h; 
\qquad
\delta^{\kappa^3}_{\bar s} h= g_{\bar s}\,\kappa^{3}\, \partial^{+n} \left[\,E \partial^{+(m-1)}h\, E^{-1} \partial^{+m}h \right]\Big|_{a^{3}},
\end{align}
where $g_{s}$ and $g_{\bar s}$ are unknown relative coefficients which are a function of $m$.
As introduced earlier, we use a convenient way of organizing derivatives with 
\begin{align}
E=e^{a \hat{\bar \partial} + b\hat{\partial}}\quad {\rm and}\quad E^{-1}=e^{-a \hat{\bar \partial} - b\hat{\partial}},\quad
\left(\hat{\bar \partial}\equiv\frac{ \bar \partial}{\partial^+}~~{\rm and}~~\hat{\partial}\equiv\frac{\partial}{\partial^+}\right).
\end{align} 

\subsection{$[ \delta_{\bar J^-}, \delta_{P^-} ]\, h$}
We here compute 
$[\, \delta_{\bar J^-} \,,\, \delta_{P^-}\,]\, h$.
To one-loop order, it has two parts
\begin{equation}
[\, \delta_{\bar J^-} \,,\, \delta_{P^-}\,]^{{\kappa^3}}\, h~=~ [\, \delta^{\kappa^3}_{\bar J^-} \,,\, \delta^{\rm free}_{P^-}\,]\, h +[\, \delta^{\rm free}_{\bar J^-} \,,\, \delta^{\kappa^3}_{P^-}\,]\, h\ .
\end{equation}
The first term of the RHS is simply
\begin{equation}\label{firstpart}
[\, \delta^{\kappa^3}_{\bar J^-} \,,\, \delta^{\rm free}_{P^-}\,]\, h ~=~ -\bar x\, [\, \delta^{\kappa^3}_{P^-} \,,\, \delta^{\rm free}_{P^-}\,]\, h\,+\, i\hat{\bar \partial} \delta^{{\kappa^3}}_{P^-} \, h\,+\, [ \delta^{\kappa^3}_{\bar s}, \delta^{\rm free}_{P^-}] h\ . 
\end{equation}
The second term can be split into three parts
\begin{align}
[\, \delta^{\rm free}_{\bar J^-} \,,\, \delta^{\kappa^3}_{P^-}\,]\, h= 
[\, i\delta_{\bar x \frac{\partial\bar\partial}{\partial^+}} \,,\, \delta^{\kappa^3}_{P^-}\,]\, h+[\, i\delta_{-x^-\bar \partial} \,,\, \delta^{\kappa^3}_{P^-}\,]\, h
+[\, i\delta_{\lambda \frac{\bar\partial}{\partial^+}} \,,\, \delta^{\kappa^3}_{P^-}\,]\, h.
\end{align}
Each of the RHS is given by
\begin{equation*}
[\, i\delta_{\bar x \frac{\partial\bar\partial}{\partial^+}} \,,\, \delta^{\kappa^3}_{P^-}\,]\, h ~=~
-\bar x [\delta^{\rm free}_{P^-}\,,\, \delta^{\kappa^3}_{P^-}] h\,-\, \kappa^3\partial^{+n} \left(2b E \partial^{+(m-1)} \delta^{\rm free}_{P^-} h E^{-1} \partial^{+m}h\right)\Big|_{a^3,\,b}\ ,
\end{equation*}
\begin{eqnarray*}
[\, i\delta_{-x^-\bar \partial} \,,\, \delta^{\kappa^3}_{P^-}\,]\, h= 
i(m-3)\delta_{\hat{\bar\partial}}\delta_{P^-}^{\kappa^3} h + i\, n\hat{\bar\partial} \delta_{P^-}^{{\kappa^3}} h +\kappa^3\partial^{+n} \left(2b E \partial^{+(m-1)} \delta^{\rm free}_{P^-} h E^{-1} \partial^{+m}h\right)\Big|_{a^{3},\,b}\ .
\end{eqnarray*}
and
\begin{equation*}
[\, i\delta_{\lambda \frac{\bar\partial}{\partial^+}} \,,\, \delta^{\kappa^3}_{P^-}\,]\, h ~=~ i\,\lambda\,\delta_{\hat{\bar\partial}}\delta_{P^-}^{{\kappa^3}} h  \, -\,i\,\lambda\,\hat{\bar\partial} \delta_{P^-}^{{\kappa^3}} h\ .
\end{equation*}
As $\big([\delta^{\rm free}_{P^-}\,,\, \delta^{\kappa^3}_{P^-}]+[\delta^{\kappa^3}_{P^-}\,,\, \delta^{\rm free}_{P^-}] \big) h=0$, one obtains
\begin{equation}\label{B10}
[\, \delta_{\bar J^-} \,,\, \delta_{P^-}\,]^{{\kappa^3}}\, h~=~  [ \delta^{{\kappa^3}}_{\bar s}, \delta^{\rm free}_{P^-}] h\,+\,i\,(m-3+\lambda)  \delta_{\hat{\bar\partial}}\delta_{P^-}^{{\kappa^3}} h \,+\,i\,(n+1-\lambda)\hat{\bar\partial} \delta_{P^-}^{{\kappa^3}} h\ ,
\end{equation}
where $\lambda({\rm helicity)}=2$. 
If we introduce a shorthand notation 
\begin{align*}\label{B11}
\Big( ~~,~~\Big) := \kappa^3 \partial^{+n} \Big(E\partial^{+m}h,\, E^{-1}\partial^{+m}h\Big)\Big|_{a^3},
\end{align*}
then we can easily distinguish the transverse derivative structure. For instance,
\begin{align*}
\delta^{\kappa^3}_{P^-} h&= \kappa^{3} \partial^{+n} \left[\,E \partial^{+m}h\, E^{-1} \partial^{+m}h \right]\Big|_{a^{3},\,b} =  2\Big(\hat\p,~\Big),\cr
\delta^{\kappa^3}_{\bar s} h&= g_{\bar s}\,\kappa^{3}\, \partial^{+n} \left[\,E \partial^{+(m-1)}h\, E^{-1} \partial^{+m}h \right]\Big|_{a^{3}} = g_{\bar s}\Big(\frac{1}{\p^+},~\Big),
\end{align*}
where we have used the antisymmetric property 
$\Big(\mathcal{O}, ~\Big)\Big|_{a^3} = \Big(~,-\mathcal{O}\Big)\Big|_{a^3}$
for a given operator $\mathcal{O}$.
Each part of the RHS of \eqref{B10} is then expressed as
\begin{align*}
[ \delta^{{\kappa^3}}_{\bar s}, \delta^{\rm free}_{P^-}] h
=-ig_{\bar s}\frac{1 }{\p^+}\left[ \Big(\hat\p,\bar\p\Big)+\Big(\hat{\bar\p},\p\Big)-
\Big(\hat{\p}\hat{\bar\p},\p^+\Big) - \Big(~,\p\hat{\bar\p}\Big)
\right],
\end{align*}
\begin{align*}
\delta_{\hat{\bar\partial}}\delta_{P^-}^{{\kappa^3}} h = &
\frac{2}{\p^+}\left[ 
\Big({\p}\hat{\bar\p},~\Big) +\Big(\hat{\p}\hat{\bar\p},\p^+\Big)
-\Big(\bar\p,\hat{\p}\Big)-\Big(\hat{\bar\p},{\p}\Big)\right],
%
\end{align*}
and
\begin{align*}
\hat{\bar\partial} \delta_{P^-}^{{\kappa^3}} h =\frac{2}{\p^+}\left[ 
\Big(\bar\p\hat\p,~ \Big) +\Big(\hat\p,\bar\p\Big)
\right].
\end{align*}
Demanding \eqref{B10} to vanish, the coefficients of each $\Big(~,~\Big)$ must vanish.
This yields two equations
\begin{align*}
2i (m+n-2) -ig_{\bar s} =0,\qquad  2i(m-1)+ig_{\bar s} =0.
\end{align*}
Together with the dimension constraint \eqref{dimconst}, one finds 
\begin{equation}\label{gsbar}
g_{\bar s}~=\,- 2 (m - 1)\ .
\end{equation}

\subsection{$[ \delta_{J^-}, \delta_{P^-} ]\, h$}
In a similar fashion, $[ \delta_{J^-}, \delta_{P^-} ]^{{\kappa^3}}\, h$ can be easily obtained, and the result is
\begin{equation}
[\, \delta_{J^-} \,,\, \delta_{P^-}\,]^{\kappa^3}\, h~=~  [ \delta^{{\kappa^3}}_{s}, \delta^{\rm free}_{P^-}]h \,+\,i\,(m-q-\lambda)  \delta_{\hat{\partial}}\delta_{P^-}^{{\kappa^3}} h \,+\,i\,(n+1+\lambda)\hat{\partial} \delta_{P^-}^{{\kappa^3}} h\ ,
\end{equation}
where $\lambda=2$ and $q$ is the power of $b$ in \eqref{1loopans}, here $q=1$.  Requiring this commutator vanish, one finds
\begin{equation}\label{gs}
g_{s}~=~ -2 (m - 3)\ .
\end{equation}

\subsection{$[ \delta_{J^-}, \delta_{\bar J^-} ]\, h$}
We want to make sure our $lc_2$ formulation is consistent with the fact that there is no counterterm at one-loop.
So far we have checked the commutation relations with dynamical Lorentz generators $[\,\delta_{\bar J^{-}}\,,\, \delta_{P^-}\,]h$ and $[\,\delta_{J^{-}}\,,\, \delta_{P^-}\,]h$ which give a nonzero contribution to one-loop order. We also need to check the commutation relation between $J^-$ and $\bar J^-$ for completeness. Hope is to see an inconsistency so that it will lead to a conclusion that one cannot construct consistent Lorentz invariant counterterms to one-loop order.

If one uses $g_{\bar s}$ \eqref{gsbar} and $g_s$ \eqref{gs}, the commutator $[ \delta_{J^-}, \delta_{\bar J^-} ]^{\kappa^3}\, h$ is simplified as
\begin{align}
&[ \delta_{J^-}, \delta_{\bar J^-} ]^{\kappa^3} h\\
&= - 2\lambda \frac{1}{\partial^+} \delta_{P^-}^{\kappa^3}h \,\bigg|_{a^3, b^1} \cr
&+i (m-\lambda)\delta_{\hat \partial} \delta^{\kappa^3}_{\bar s}h + i (n+\lambda)\hat{\partial}\delta^{\kappa^3}_{\bar s} h -g_{\bar s}\kappa^3
\partial^{+n} \left( E \hat \partial\partial^{+(m-1)} h \, E^{-1} \partial^{+m}h\right) \bigg|_{a^3}\cr
&-i (m +\lambda-2) \delta_{\hat{\bar\partial}}\delta^{\kappa^3}_{s} h + i (-n + \lambda) \hat{\bar\partial} \delta^{\kappa^3}_{s}h  +g_s \kappa^3
\partial^{+n} \left( E\hat{\bar\partial}\partial^{+(m-1)} h \, E^{-1} \partial^{+m} h \right)\bigg|_{a^2,b^1}.\nonumber
\end{align}
It is straightforward to see that this commutator vanishes for any value of $m$.\\

We summarize the calculations.
We started with an Ansatz  \eqref{1loopans} which is expressed in a coherent-state like form which satisfies all the commutation relations with the kinematical generators. From dimensional grounds, we have a constraint  \eqref{dimconst}, $n+2m =3$, 
that relates two unknown parameters of the Ansatz. The commutators of Hamiltonian transformation with dynamical boosts lead to two other equations
$g_s =-2 (m - 3)$, and $g_{\bar s} = -2(m-1)$
We found that the commutator between two dynamical boost transformations 
$[\delta_{J^-},\delta_{\bar J^-}]h$ is consistent with the obtained $g_s$ and $g_{\bar s}$.  We conclude that   $[\delta_{J^-},\delta_{\bar{J}^-}]h$ does not determine $m$, and thus the unknown powers of the light-cone derivatives $\partial^+$ is not determined by closing the Lorentz algebra, yielding infinitely many counterterms.


\section{Two loop calculations}\label{barhbarh}
\noindent$\bullet$ $\delta^{{\kappa^5}}_{P^-} h~\sim~  {\partial}^6\bar h\, \bar h\ $\\
For this type, one easily sees that there is no possible Ansatz with two $\bar h$'s for $\delta^{g_2}_s h$,
due to helicity and the numbers of allowed transverse derivatives. The other Ans\"{a}tze are of a similar form as shown in the previous type
\begin{eqnarray}
\delta^{{\kappa^5}}_{P^-} h&=&\kappa^{5} \partial^{+n} \left[\,E \partial^{+m}\bar h\, E^{-1} \partial^{+m}\bar h \right]\Big|_{b^6} \ ,\\
\delta^{{\kappa^5}}_{\bar s} h&=& g_{\bar s}\,\kappa^{5}\, \partial^{+n} \left[\,E \partial^{+(m-1)}\bar h\, E^{-1} \partial^{+m}\bar h \right]\Big|_{b^5} .
\end{eqnarray}
Straightforward calculations lead to 
\begin{align}\label{2loopeq}
[\delta_{\bar J^-}, \delta_{P^-}]^{{\kappa^5}}h = 
[\delta^{{\kappa^5}}_{\bar s}, \delta^0_{P^-}]h +i(m-2)\delta_{\hat{\bar\partial}}\delta^{{\kappa^5}}_{P^-}h + i (n-1)\hat{\bar\partial}\delta^{{\kappa^5}}_{P^-} h ,
\end{align}
where $n+2m= 5$.  
The first term of the RHS of \eqref{2loopeq} is expressed as
\begin{align}
[\delta^{{\kappa^5}}_{\bar s}, \delta^0_{P^-}]h &= -ig_{\bar s}
\kappa^5 \partial^{+(n-1)}
\Big[\,E \partial^{+m}\hat{\partial}\bar h\, E^{-1} \partial^{+(m+1)}\hat{\bar\partial}\bar h
- E \partial^{+(m+1)}\hat{\partial}\bar h\, E^{-1} \partial^{+m}\hat{\bar\partial}\bar h\cr
&\qquad\qquad\qquad\quad -E \partial^{+m}\hat{\partial}\hat{\bar\partial}\bar h\, E^{-1} \partial^{+(m+1)}\bar h
+E \partial^{+(m+1)}\hat{\partial}\hat{\bar\partial}\bar h\, E^{-1} \partial^{+m}\bar h
 \Big]\Big|_{b^5}.\nn
\end{align}
We then have
\begin{align}
[\delta_{\bar J^-}, \delta_{P^-}]^{{\kappa^5}}h 
&= \big( 2i (m-2)+ 2i(n-1) -ig_{\bar s}\big) \big[~ \hat{\partial}\,,\, \hat{\bar \partial}\partial^+~\big]\cr
&+\big( 2i (m-2)+ig_{\bar s}\big) \big[~ \hat{\partial}\partial^+\,,\, \hat{\bar \partial}~\big]\cr
&+\big( 2i (m-2)+ig_{\bar s}\big) \big[~ \hat{\partial}\hat{\bar \partial}\,,\, \partial^+\big]\cr
&+\big( 2i (m-2)+ 2i(n-1) -ig_{\bar s}\big) \big[~ \hat{\partial}\hat{\bar \partial}\partial^+\,,\, ~\big], \nn
\end{align}
where 
$ \big[~ \,,\, ~\big] \equiv \kappa^5 \partial^{+(n-1)}
\Big[\,E \partial^{+m}\bar  h, E^{-1} \partial^{+m}\hat{\bar\partial}\bar h\Big]_{b^5}$.  By setting this equation zero,  we get, for $ n=5-2m$,
\begin{equation} 
g_{\bar s}~=~ -2\, (m - 2) 
\end{equation}

It is crucial that because of helicity structure, the spin part of $\delta_{J^-}h$ does not exist in this case.
With this fact, we can further restrict the value of $m$:
\begin{align}
[\delta_{J^-}, \delta_{P^-}]^{{\kappa^5}}h &=[\delta^{\kappa^5}_{J^-}, \delta^0_{P^-}]h +[\delta^0_{J^-}, \delta^{{\kappa^5}}_{P^-}]h,\nn
\end{align}
where
\begin{align}
[\delta^{{\kappa^5}}_{J^-}, \delta^0_{P^-}]h&= +i\hat{\partial} \delta^{\kappa^5}_{P^-} h+ i x [\delta^{{\kappa^5}}_{P^-},\delta^{0}_{P^-}]h,\cr
[\delta^0_{J^-}, \delta^{{\kappa^5}}_{P^-}]h&=[\delta_{ix{P^-}^0}, \delta^{{\kappa^5}}_{P^-}]h + [\delta_{-ix^- \partial}, \delta^{{\kappa^5}}_{P^-}]h +[\delta_{-i\lambda \hat{\partial}}, \delta^{{\kappa^5}}_{P^-}]h, \cr
[\delta_{ix{P^-}^0}, \delta^{{\kappa^5}}_{P^-}]h&=i x [\delta^0_{P^-},\delta^{{\kappa^5}}_{P^-}]h,\cr
[\delta_{-ix^- \partial}, \delta^{{\kappa^5}}_{P^-}]h&=
im\delta_{\hat{\partial}}\delta^{{\kappa^5}}_{P^-}h + i n\hat{\partial}\delta^{{\kappa^5}}_{P^-} h\cr
&\quad
-i\kappa^5 \partial^{+n} (bE\partial^{+m}\hat{\partial}^2 \bar h \,
E^{-1}\partial^{+m} \bar h  -b E\partial^{+m} \bar h \,
E^{-1}\partial^{+m} \hat{\partial}^2\bar h ),\cr
[\delta_{-2i \hat{\partial}}, \delta^{{\kappa^5}}_{P^-}]h&=
2i\delta_{\hat{\partial}}\delta^{{\kappa^5}}_{P^-}h +2i\hat{\partial}\delta^{\kappa^5}_{P^-} h .\nn
\end{align}
Using the identity \eqref{A14}, we find that 
\begin{align}
[\delta_{J^-}, \delta_{P^-}]^{{\kappa^5}}h=
i(m-4)\delta_{\hat{\partial}}\delta^{{\kappa^5}}_{P^-}h + i (n+3)\hat{\partial}\delta^{{\kappa^5}}_{P^-} h ,
\end{align}
which yields
\begin{align}
m = 4, \qquad  n = -3.
\end{align}

We now see that these solutions do satisfy the dimension constraint $2m+n=2l+1$. Here two-loop means $l=2$ and thus
\begin{align*}
2m + n = -6 +11 = 5.
\end{align*}
Hence we found a unique solution for the two loop counterterm  
\begin{align}
m=4, \qquad n=-3, \qquad g_{\bar s} = -4.
\end{align}
Thus we have
\begin{align}
\delta^{{\kappa^5}}_{P^-} h&=\kappa^{5}\frac{1}{\partial^{+3}} \left[\,E \partial^{+4}\bar h\, E^{-1} \partial^{+4}\bar h \right]\Big|_{b^6} \ ,\\
\delta^{{\kappa^5}}_{\bar s} h&= -4\,\kappa^{5}\,\frac{1}{\partial^{+3}} \left[\,E \partial^{+3}\bar h\, E^{-1} \partial^{+4}\bar h \right]\Big|_{b^5} .
\end{align}\\

%
\providecommand{\href}[2]{#2}\begingroup\raggedright\endgroup
\end{document}